\documentclass[prl,twocolumn,floatfix]{revtex4}

\usepackage{amssymb}
\usepackage{amsmath}
\usepackage{epsfig}

\begin{document}

\title{Why temperature chaos in spin glasses is hard to observe}
\author{T. Aspelmeier}
\author{A. J. Bray}
\author{M. A. Moore}
\affiliation{Department of Physics and Astronomy,
University of Manchester, Manchester M13 9PL, UK}

\begin{abstract}
The overlap length of a three-dimensional Ising spin glass on a cubic
lattice with Gaussian interactions has been estimated numerically by
transfer matrix methods and within a Migdal-Kadanoff renormalization
group scheme. We find that the overlap length is large, explaining why it
has been difficult to observe spin glass chaos in numerical simulations and
experiment.
\end{abstract} 
\maketitle

Chaos, rejuvenation, memory, and aging in spin glasses are currently
being intensively studied
\cite{JonasonEtAl98,BouchaudEtAl01,Rizzo01,YoshinoEtAl01,BerthierBouchaud02,%
KrzakalaMartin02,JonssonEtAl02,TakayamaHukushima02,BilloireMarinari00}.
``Chaos'' refers to the property that equilibrium states in the
ordered phase of spin glasses are sensitive to arbitrarily small
changes in the couplings or in temperature, and is one possible
explanation for memory effects in spin glasses \cite{JonasonEtAl98},
although other mechanisms are also present
\cite{BerthierBouchaud02}. There is evidence both
for and against temperature chaos in computer simulations: some
authors claim to see the effects of chaos \cite{TakayamaHukushima02},
others have failed to see it
\cite{BilloireMarinari00,BilloireMarinari02}.
In this Letter we start from the droplet \cite{FisherHuse86} 
or scaling \cite{BrayMoore87} pictures and calculate numerically the 
overlap length $L^*(T,\Delta T)$, i.e.\ the length scale beyond which 
spins at temperatures $T$ and $T+\Delta T$ become
uncorrelated with each other, 
for a three-dimensional Edwards-Anderson Ising spin
glass and also within a Migdal-Kadanoff renormalization group scheme
(MKRG), both with a Gaussian distribution of couplings with unit
variance. We show that this length scale is large within much of the 
parameter space, and only just comes down to magnitudes accessible
to numerical simulations for certain values of $T$ and $\Delta T$. 
We believe that this is why some workers have been unable to see 
chaos in their investigations. As a by-product, we  will also
find the root mean square droplet interface free energy
$F(T)\equiv\sqrt{\langle F_{\text{int}}^2(T)\rangle}$ and entropy
$S(T)\equiv\sqrt{\langle S_{\text{int}}^2(T)\rangle}$. The angle
brackets indicate averaging over realizations of the bond
couplings. We show that $S(T)\sim\sqrt{T}$ for small $T$, contrary to
previous arguments \cite{HuseFisher86}. 


In the MKRG scheme for three
dimensions, renormalized bonds after $n$ renormalization steps
$J^{(n)}$ are obtained from the set of bonds $\{J^{(n-1)}\}$ after
$n-1$ renormalization steps using the relation
\begin{equation}
\frac{J^{(n)}}{T} = 
\sum_{i=1}^{4}\tanh^{-1}\left(
\tanh\frac{J_{1i}^{(n-1)}}{T}\tanh\frac{J_{2i}^{(n-1)}}{T}\right),
\end{equation}
where $J_{1,2\,i}^{(n-1)}$ are randomly drawn members of the bond pool
(``pool method''). For a detailed description of the method in the
present context we refer the reader to \cite{BanavarBray87}. The
renormalized bonds play the role of the interface free energy of a
system of linear size $L=2^n$, to wit $\frac 12 F(T)=\langle
(J^{(n)})^2\rangle^{1/2}$, the angle brackets indicating the average
over the bond pool.  Similarly, the interface entropy is given by
$\frac 12 S(T) = \lim_{\delta T\to 0}\langle (J^{(n)} - {J'}^{(n)})^2
\rangle^{1/2}/\delta T$, where ${J'}^{(n)}$ is the corresponding
member of a bond pool which has been evolved at temperature $T+\delta
T$.

The MKRG applied to chaos in spin glasses has the advantages that 
large length scales are easily accessible numerically and that it is
possible to access and estimate the overlap length directly.  Banavar
and Bray \cite{BanavarBray87} have shown that the overlap length $L^*$
is related to the ratio of the interface entropy and interface free
energy. Introducing the usual generalized stiffness coefficients
$\Upsilon(T)$ and $\sigma(T)$ via
\begin{equation}
\label{scalingeq}
F(T) = \Upsilon(T)L^{\theta} \text{ and }
S(T) = \sigma(T)L^{d_s/2},
\end{equation}
(valid for large system sizes $L$), a measure of the overlap length 
for a temperature step $\Delta T \ll T$ is obtained by equating $F(T)$ 
and $S(T)\,\Delta T$:  
\begin{equation}
\label{lstareq}
L^* = \left(\frac{\Upsilon}{\sigma\Delta T}\right)^{1/\zeta}. 
\end{equation}
Here, $\zeta=\frac{d_s}{2}-\theta$ is the chaos exponent, where $d_s$ 
is the fractal dimension of the droplet interface and $\theta$ is the 
droplet excitation energy exponent.  

Independently, the overlap length can be obtained as the length scale
on which two initially identical bond pools $\{J\}$ and $\{J'\}$,
evolved at two different temperatures, decorrelate. A measure for this
is $L^*=2^{n_0}$ where $n_0$ is such that
\begin{equation}
\frac{\left\langle\left(J^{(n_0)}-{J'}^{(n_0)}\right)^2\right\rangle}
{\langle(J^{(n_0)})^2\rangle + \langle({J'}^{(n_0)})^2\rangle} = \frac 12.
\label{direct}
\end{equation}
This method does not require a small temperature difference. When the
temperature difference does happen to be small, however, it agrees with
the previous method up to a numerical factor close to unity (see below).

\begin{figure}
\epsfig{file=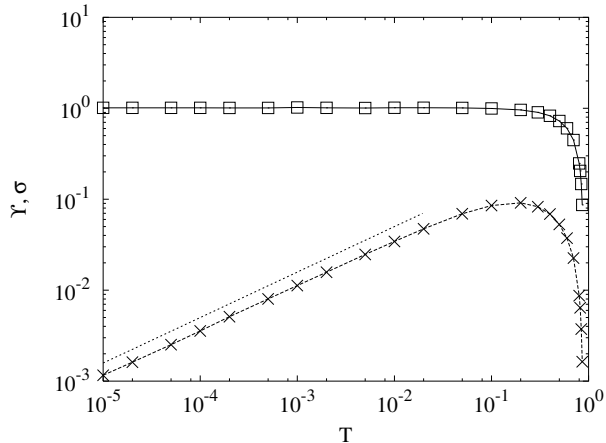,width=\columnwidth}\\
\vspace*{-0.3cm}
\caption{Data for $\Upsilon(T)$ (solid line) and $\sigma(T)$ (dashed line)
from Migdal-Kadanoff renormalization. The dotted line has slope 1/2 for 
comparison.}
\label{upssig}
\end{figure}

Repeating the numerical work from \cite{BanavarBray87} with greater
accuracy (pool size 2000000 and `long double' precision) and for a
larger set of temperatures, we find $\Upsilon(T)$ and $\sigma(T)$ as
shown in Fig.~\ref{upssig}.
The data for $\sigma(T)$ in this figure has been obtained using a
fixed \textit{relative} temperature change $\epsilon = \delta T/T =
10^{-6}$ at each temperature.

Huse and Fisher \cite{HuseFisher86} argued that a droplet interface
can be regarded as a collection of roughly independent two level
systems, each of which has on average an entropy proportional to $T$,
which should give rise to $S(T)\sim T$. This is clearly violated by
the data in Fig.~\ref{upssig}. We will come back to this point later.

\begin{figure}[bt]
\epsfig{file=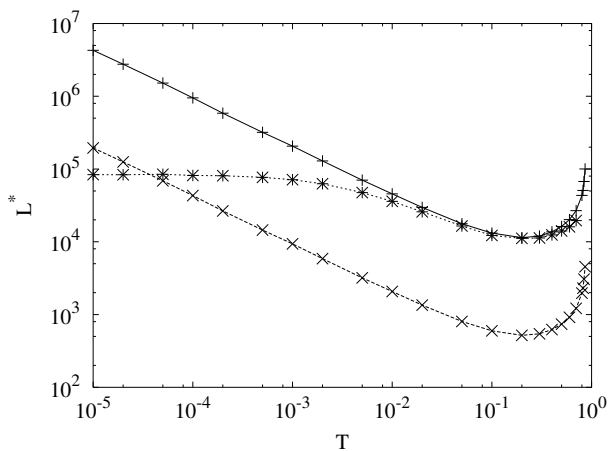,width=\columnwidth}\\
\vspace*{-0.3cm}
\caption{Plot of the overlap length $L^*$ as determined from the MKRG
  data and Eq.~\eqref{lstareq} with $\zeta=0.7448$ for $\Delta T=0.01$ (solid
  line) and $\Delta T=0.1$ (dashed line). The dotted line shows a direct
  determination of $L^*$ for fixed absolute temperature shift $\Delta T=0.01$,
  (see text).}
\label{lstar}
\end{figure}

The overlap length $L^*$ as obtained from the data for $\Upsilon$ and
$\sigma$ using Eq.~\eqref{lstareq} is shown in Fig.~\ref{lstar}.
This requires knowing $\zeta$. For the $d$-dimensional
Migdal-Kadanoff scheme, $d_s=d-1$ (i.e.\ $d_s=2$ in the present case),
and $\theta$ was estimated from the numerics to be $\theta= 0.25519\pm
0.00005$ for $d=3$, therefore $\zeta\approx 0.7448$. For illustration,
the temperature shift has been arbitrarily set to $\Delta T=0.01$ and
$\Delta T=0.1$ for this plot. Since for these values of the parameters
the temperature shift $\Delta T$ is not much smaller than $T$ for
small $T$, Eq.~\ref{lstareq} is not expected to hold and the ``true''
overlap length differs from the data shown. A direct determination of
the overlap length (as the decorrelation length of the two bond pools
as explained above) with $\Delta T=0.01$ is shown in Fig.~\ref{lstar}
for comparison. A small constant has been added to the latter curve to 
make the two ways of determining $L^*$ agree at the minimum. (This   
adjustment reflects the arbitrariness of the constant 1/2 in the 
definition, through Eq.\ (\ref{direct}), of $L^*$.)\   With this 
correction, the two curves for $\Delta T=0.01$ agree very well in a 
region around the minimum; they differ for small $T$ since 
$\Delta T \ge T$ and for $T$ close to the critical temperature because 
of increasing influence of critical point controlled decorrelation
\cite{NifleHilhorst92}.

\begin{figure}
\hspace*{-0.3\columnwidth}
\vspace*{-0.27\columnwidth}\\
\hspace*{-0.35\columnwidth}
\epsfig{file=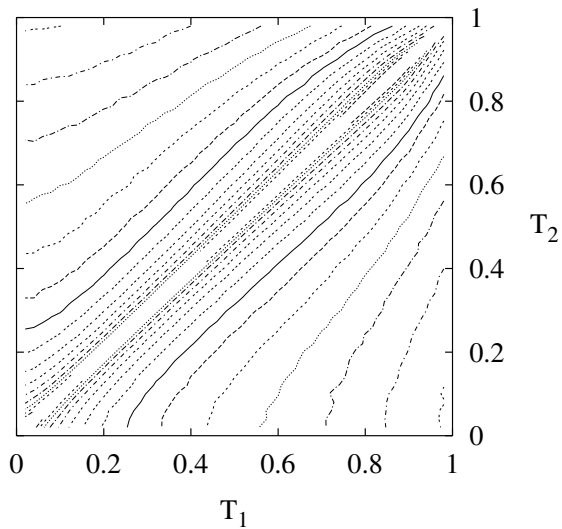,width=1.3\columnwidth}
\vspace*{-0.25\columnwidth}
\caption{The overlap length $L^*$ as a function of two temperatures 
$T_1$ and $T_2$.  The contour lines are situated at $L^*=10^{n/5}$,
beginning with $n=8$ (i.e.\ $L^*\approx 40$ lattice spacings) 
and counting from the
upper left and lower right hand corners towards the middle diagonal,
where $L^*$ diverges. The highest contour line shown is at $n=22$ 
where $L^*\approx 25000$.}
\label{lstarplane}
\end{figure}

Fig.~\ref{lstarplane} shows a contour plot of the overlap length in the
two-temperature plane, obtained from MKRG in the same manner as before.
This figure shows that the overlap length is nowhere less than 40 for
$T_1,T_2<T_c=0.89645\dots\,$. Note that in agreement with
\cite{NifleHilhorst92} the overlap length is nonzero even above $T_c$.

Obviously, $L^*$ is extraordinarily large for most values of $T$ and
never comes into a numerically accessible range, which, by today's
standards, would be around $L^*\approx 20$.

While the Migdal-Kadanoff renormalization method allows study of
large length scales, it has drawbacks: the fractal dimension of a 
domain wall $d_s$ on a cubic lattice is not well approximated, 
so the structure of an interface is significantly
different on a hierarchical Berker lattice \cite{BerkerOstlund79} (for
which MKRG is exact) than on a cubic lattice. It is therefore not
obvious that the overlap length as obtained above is characteristic of
a three-dimensional Edwards-Anderson spin glass as well. In order to
test this, we used the transfer matrix method as described in
\cite{BrayMoore85} to numerically obtain the interface free energy and
entropy of small spin glass samples. This method has the disadvantage
that due to demands on computer time and memory, only samples of size
up to $L=4$ are accessible with reasonable statistics. On the positive
side, there are no approximations involved whatsoever. This approach
is therefore complimentary to the Migdal-Kadanoff scheme.
We shall use Eq.~\eqref{lstareq}, which applies generally wherever the droplet
picture holds,
to provide estimates of
$L^*$ from measurements of $\Upsilon$ and $\sigma$.

In order to obtain the interface free energy $F_{\text{int}}$ of a
spin glass sample of size $L\times L\times (L+1)$ at temperature $T$,
first the free energy of the sample with periodic boundary conditions
along the first two dimensions and fixed boundary conditions ($+1$ on
both sides) along the third dimension is calculated. Next, the free
energy of the same sample with boundary condition $+1$ on the one side
and $-1$ on the other is calculated. The difference is the interface
free energy. The interface entropy $S_{\text{int}}$ is obtained by
repeating the above for temperature $T+\delta T$ and taking the
numerical derivative of $F_{\text{int}}$. From these data, the root
mean square averages $F(T)$ and $S(T)$ are calculated.

\begin{figure}
\epsfig{file=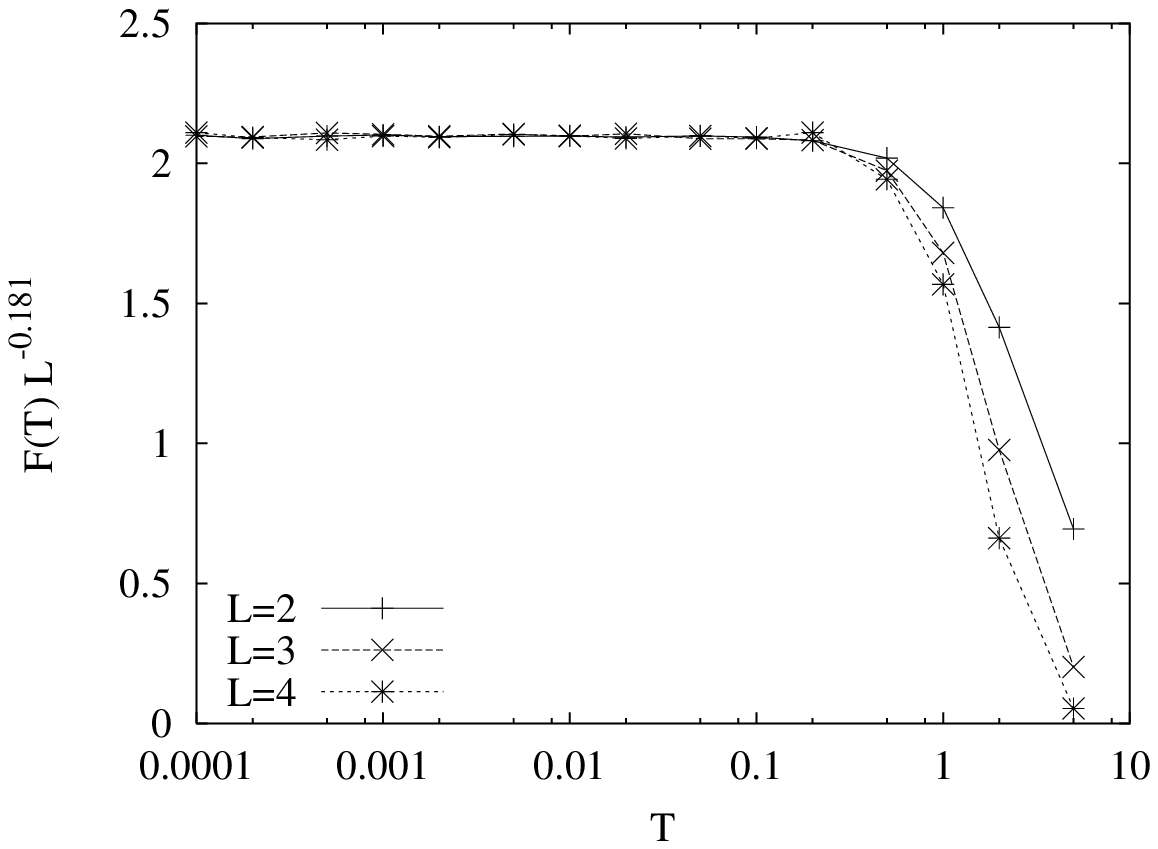,width=\columnwidth}
\epsfig{file=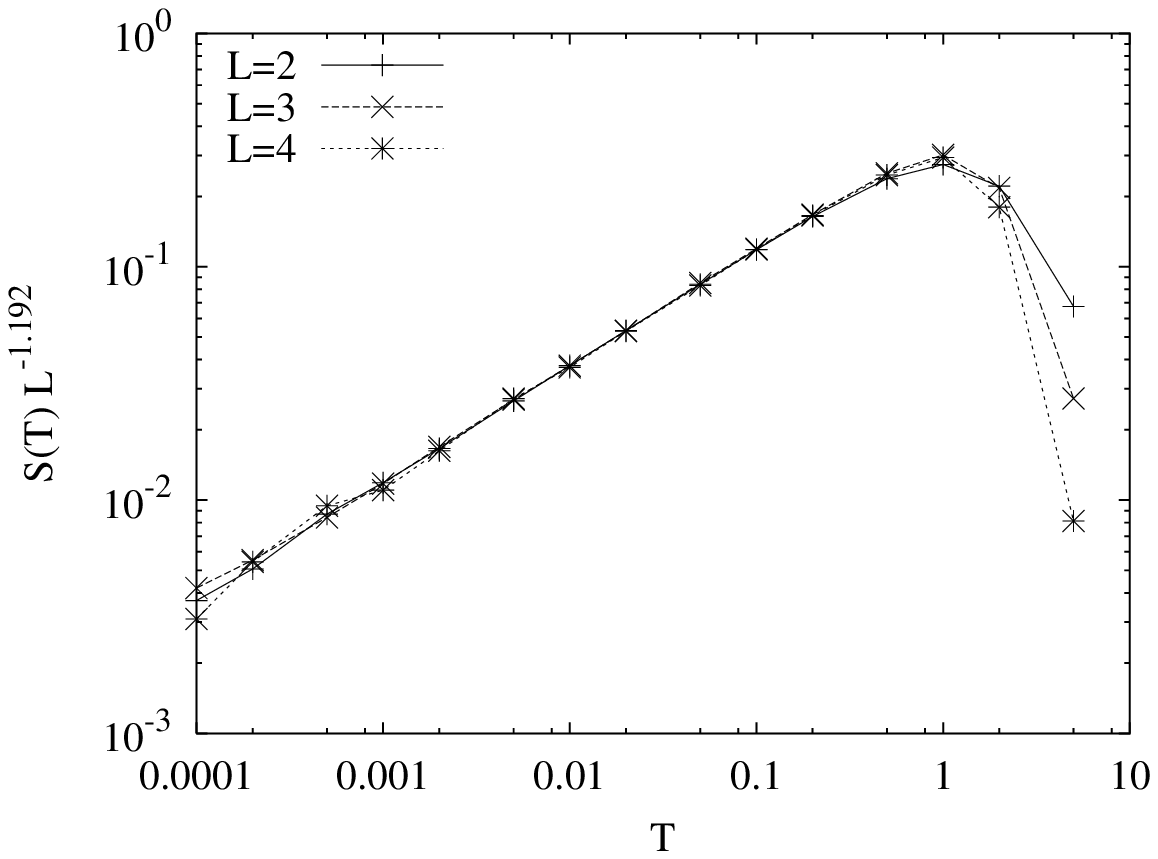,width=\columnwidth}\\
\vspace*{-0.3cm}
\caption{Scaling plot for the interface free energy (top) and entropy 
(bottom). The numbers of samples used were $2\cdot 10^5$ ($L=2$),
$1\cdot 10^5$ ($L=3$), and $2\cdot 10^4$ ($L=4$). With $\theta=0.181$
and $d_s/2=1.192$ good data collapse is achieved.}
\label{scaling}
\end{figure}

The data for the interface free energy and entropy are shown in
Fig.~\ref{scaling} as scaling plots for system sizes $L=2,3,4$. The
entropy has been obtained with a fixed relative temperature shift
$\epsilon=\delta T/T = 10^{-4}$.
The plot for $F(T)$ allows for an estimate of $\theta$, the one for
$S(T)$ for an estimate of $d_s$ (cf.\ Eq.~\eqref{scalingeq}). The
values obtained are $\theta\approx 0.18$ and $d_s\approx 2.38$,
leading to $\zeta\approx 1.01$. While these numbers are merely crude
estimates (corrections to scaling are expected to be significant for
these small system sizes) and therefore no error bars have been
supplied, they certainly lie in the expected range and the value of
$d_s$, in particular, shows that the droplet interface structure is
much better captured even for these small systems than in the
MKRG. The scaling in Fig.~\ref{scaling} naturally breaks down close to
the critical temperature, due to the small system sizes.

\begin{figure}
\epsfig{file=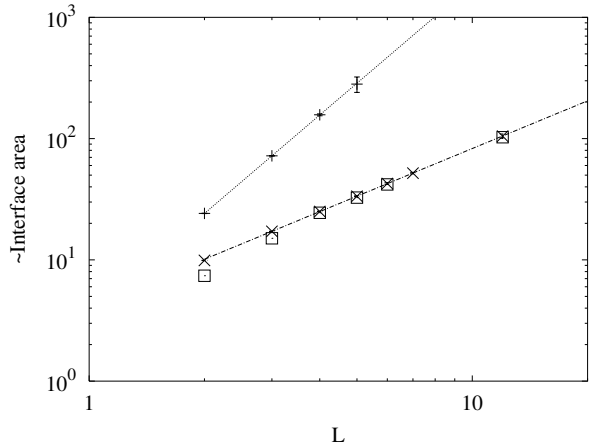,width=\columnwidth}\\
\vspace*{-0.3cm}
\caption{Estimate of the interface area from a bond perturbation calculation.
The + symbols are for $d=3$, and the dashed straight line is a best
fit with slope $d_s=2.697\pm 0.002$. The numbers of samples used are
as in Fig.~\ref{scaling} and $100$ for $L=5$.  The crosses are for
$d=2$ and the dotted straight line is a fit with $d_s=1.3$ (the line
is to guide the eye only; better data can be found in
\cite{BrayMoore87}). The squares represent the average of
$S^2(T)/T$ in $d=2$ over a temperature range where this quantity
is approximately constant, multiplied by an arbitrary number in
order to make it comparable to the other data. }
\label{Lint}
\end{figure}

The fractal dimension $d_s$ can also be estimated at zero temperature
from a bond perturbation calculation as in \cite{BrayMoore87}. The
result, shown in Fig.~\ref{Lint}, is $d_s\approx 2.7$, obtained from sample
sizes $L=2,3,4,5$.
The discrepancy between this result and the one above illustrates the
influence of finite size effects. A similar comparison in two
dimensions \footnote{Since $T_c=0$ in two dimensions, the entropy
scaling only works if $L$ is much smaller than the correlation length,
i.e.\ only at small temperatures.}, also shown in Fig.~\ref{Lint},
gives the same kind of discrepancy for small $L$, which however is
removed when going to larger $L$. It is found that the value obtained
from bond perturbation is more reliable than the one obtained from a
scaling plot of the entropy; the bond perturbation data lie almost
perfectly on a straight line even down to $L=2$. In three dimensions,
the same seems to be true since $d_s\approx 2.7$ compares well with
$d_s\approx 2.68$ from
\cite{PalassiniYoung99}.

The interface entropy in Fig.~\ref{scaling} again scales as $S(T)\sim
\sqrt{T}$, showing that this behavior is not special to the MKRG. In order to
understand this, we return to the two-level-system argument of
\cite{HuseFisher86}. The interface entropy is calculated from the change in
entropy after a change of boundary conditions, therefore it is
necessary to know the excitation energy of each two level system
\textit{under both boundary conditions}. That these will in general be
different becomes apparent when considering, e.g., a single spin on
the interface: since it is on the interface, at least one but not all
of its neighboring spins change sign when changing boundary
conditions, thus giving rise to a different effective field for the
spin, i.e.\ a different excitation energy for flipping this spin.
Therefore, the interface entropy contribution from a two-level system
is in fact $\Delta S(\Delta^+, \Delta^-)=S_{\text{2-lev}}(\Delta^+) -
S_{\text{2-lev}}(\Delta^-)$, where $\Delta^{\pm}$ are the excitation
energies under the two boundary conditions. It is easy to check that
the entropy of a two-level system is
\begin{equation}
S_{\text{2-lev}}(\Delta) = 
\log\left(2\cosh\frac{\Delta}{2T}\right) - 
\frac{\Delta}{2T}\tanh\frac{\Delta}{2T}.
\end{equation}
The function $S_{\text{2-lev}}(\Delta)$ has a maximum at $\Delta=0$
and decays to zero on scale $T$, therefore $\Delta
S(\Delta^+,\Delta^-)$ is non-zero essentially only in two perpendicular
strips of width $T$ along the $\Delta^{\pm}$ axes, excluding their
overlap region around the origin. Thus only those two-level systems
contribute to the interface entropy which have excitation energies
$\Delta^+ < T$
\textit{and} $\Delta^- > T$ (or vice versa). This implies that the
second moment, taken with the joint probability distribution
$p(\Delta^+,\Delta^-)$, goes as
\begin{equation}
\langle \Delta S^2\rangle \sim
 \int_T^{\infty}d\Delta^+\int_0^T d\Delta^- p(\Delta^+,\Delta^-)
\sim T, \quad T\to 0, \nonumber
\end{equation}
provided the marginal distribution $\int_0^{\infty}d\Delta^+
p(\Delta^+,\Delta^-)$ is nonzero for $\Delta^-=0$. This result is very
well supported by the data in Figs.~\ref{upssig} and
\ref{scaling}.

\begin{figure}[tb]
\epsfig{file=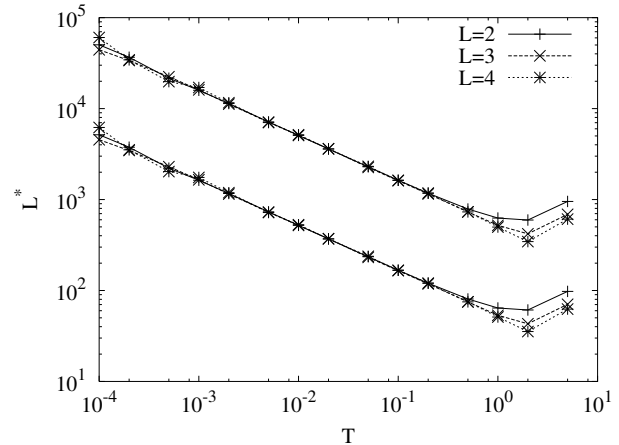,width=\columnwidth}\\
\vspace*{-0.3cm}
\caption{Overlap length as calculated from the transfer matrix method. 
Top curves are for $\Delta T=0.01$, bottom curves for $\Delta
T=0.1$. For this plot, $\zeta=1.01$ has been used as obtained from
Fig.~\ref{scaling}. There is no qualitative difference, however, if a
more realistic $\zeta=1.15$ is used.}
\label{lstartrans}
\end{figure}

The overlap length as estimated from $F(T)$, $S(T)$, and
Eq.~\eqref{lstareq} is shown in Fig.~\ref{lstartrans}.
While being smaller than for the MKRG, $L^*$ is still very large
($L^*>30$ for $\Delta T < 0.1$) and thus (just) out of range of
numerical simulations.  In fact with the parameters as used in
\cite{TakayamaHukushima02} ($T_1=0.7$, $T_2=0.4$), the overlap 
length is estimated to be $L^*\approx 20$, while the authors are using
a system of size $L=24$. This is in accord with the authors' claim that
they have been able to observe the effects of temperature chaos.
Again, we would expect modifications to the small temperature behavior 
of $L^*$ if a fixed absolute temperature shift is used 
(cf.\ Fig.~\ref{lstar}), but not in the region of the minimum.

Finally it is worth remarking that the large value of the overlap length
will have implications also for experiments on memory and rejuvenation. 
The effects of chaos are only visible when the length scale $L(t)$ within
which the spins are well-equilibrated after waiting for a time $t$, 
is larger than the overlap length $L^{*}$. As $L(t)$ only increases 
very slowly with time, many experiments will not satisfy the criterion 
$L(t)> L^{*}$ and it will then be inappropriate to use chaos ideas to 
explain what is happening.

\begin{acknowledgments}
T.A.\ acknowledges support by the German Academic Exchange Service
(DAAD) under a postdoc fellowship.
\end{acknowledgments}

\bibliography{Spinglass}

\begin{thebibliography}{18}
\expandafter\ifx\csname natexlab\endcsname\relax\def\natexlab#1{#1}\fi
\expandafter\ifx\csname bibnamefont\endcsname\relax
  \def\bibnamefont#1{#1}\fi
\expandafter\ifx\csname bibfnamefont\endcsname\relax
  \def\bibfnamefont#1{#1}\fi
\expandafter\ifx\csname citenamefont\endcsname\relax
  \def\citenamefont#1{#1}\fi
\expandafter\ifx\csname url\endcsname\relax
  \def\url#1{\texttt{#1}}\fi
\expandafter\ifx\csname urlprefix\endcsname\relax\def\urlprefix{URL }\fi
\providecommand{\bibinfo}[2]{#2}
\providecommand{\eprint}[2][]{\url{#2}}

\bibitem[{\citenamefont{Bouchaud et~al.}(2001)\citenamefont{Bouchaud, Dupuis,
  Hammann, and Vincent}}]{BouchaudEtAl01}
\bibinfo{author}{\bibfnamefont{J.-P.} \bibnamefont{Bouchaud}},
  \bibinfo{author}{\bibfnamefont{V.}~\bibnamefont{Dupuis}},
  \bibinfo{author}{\bibfnamefont{J.}~\bibnamefont{Hammann}}, \bibnamefont{and}
  \bibinfo{author}{\bibfnamefont{E.}~\bibnamefont{Vincent}},
  \bibinfo{journal}{Phys. Rev. B} \textbf{\bibinfo{volume}{65}},
  \bibinfo{pages}{024439} (\bibinfo{year}{2001}).

\bibitem[{\citenamefont{Rizzo}(2001)}]{Rizzo01}
\bibinfo{author}{\bibfnamefont{T.}~\bibnamefont{Rizzo}}, \bibinfo{journal}{J.
  Phys. A} \textbf{\bibinfo{volume}{34}}, \bibinfo{pages}{5531}
  (\bibinfo{year}{2001}).

\bibitem[{\citenamefont{Yoshino et~al.}(2001)\citenamefont{Yoshino,
  Lema\^{\i}tre, and Bouchaud}}]{YoshinoEtAl01}
\bibinfo{author}{\bibfnamefont{H.}~\bibnamefont{Yoshino}},
  \bibinfo{author}{\bibfnamefont{A.}~\bibnamefont{Lema\^{\i}tre}},
  \bibnamefont{and} \bibinfo{author}{\bibfnamefont{J.-P.}
  \bibnamefont{Bouchaud}}, \bibinfo{journal}{Eur. Phys. J. B}
  \textbf{\bibinfo{volume}{20}}, \bibinfo{pages}{367} (\bibinfo{year}{2001}).

\bibitem[{\citenamefont{Berthier and Bouchaud}(2002)}]{BerthierBouchaud02}
\bibinfo{author}{\bibfnamefont{L.}~\bibnamefont{Berthier}} \bibnamefont{and}
  \bibinfo{author}{\bibfnamefont{J.-P.} \bibnamefont{Bouchaud}}
  (\bibinfo{year}{2002}), \eprint{cond-mat/0202069}.

\bibitem[{\citenamefont{Krzakala and Martin}(2002)}]{KrzakalaMartin02}
\bibinfo{author}{\bibfnamefont{F.}~\bibnamefont{Krzakala}} \bibnamefont{and}
  \bibinfo{author}{\bibfnamefont{O.~C.} \bibnamefont{Martin}}
  (\bibinfo{year}{2002}), \eprint{cond-mat/203499}.

\bibitem[{\citenamefont{J{\"o}nsson et~al.}(2002)\citenamefont{J{\"o}nsson,
  Yoshino, and Nordblad}}]{JonssonEtAl02}
\bibinfo{author}{\bibfnamefont{P.~E.} \bibnamefont{J{\"o}nsson}},
  \bibinfo{author}{\bibfnamefont{H.}~\bibnamefont{Yoshino}}, \bibnamefont{and}
  \bibinfo{author}{\bibfnamefont{P.}~\bibnamefont{Nordblad}}
  (\bibinfo{year}{2002}), \eprint{cond-mat/0203444}.

\bibitem[{\citenamefont{Takayama and Hukushima}(2002)}]{TakayamaHukushima02}
\bibinfo{author}{\bibfnamefont{H.}~\bibnamefont{Takayama}} \bibnamefont{and}
  \bibinfo{author}{\bibfnamefont{K.}~\bibnamefont{Hukushima}}
  (\bibinfo{year}{2002}), \eprint{cond-mat/0205276}.

\bibitem[{\citenamefont{Jonason et~al.}(1998)\citenamefont{Jonason, Vincent,
  Hammann, Bouchaud, and Nordblad}}]{JonasonEtAl98}
\bibinfo{author}{\bibfnamefont{K.}~\bibnamefont{Jonason}},
  \bibinfo{author}{\bibfnamefont{E.}~\bibnamefont{Vincent}},
  \bibinfo{author}{\bibfnamefont{J.}~\bibnamefont{Hammann}},
  \bibinfo{author}{\bibfnamefont{J.-P.} \bibnamefont{Bouchaud}},
  \bibnamefont{and} \bibinfo{author}{\bibfnamefont{P.}~\bibnamefont{Nordblad}},
  \bibinfo{journal}{Phys. Rev. Lett.} \textbf{\bibinfo{volume}{81}},
  \bibinfo{pages}{3243} (\bibinfo{year}{1998}).

\bibitem[{\citenamefont{Billoire and Marinari}(2000)}]{BilloireMarinari00}
\bibinfo{author}{\bibfnamefont{A.}~\bibnamefont{Billoire}} \bibnamefont{and}
  \bibinfo{author}{\bibfnamefont{E.}~\bibnamefont{Marinari}},
  \bibinfo{journal}{J. Phys. A} \textbf{\bibinfo{volume}{33}},
  \bibinfo{pages}{L265} (\bibinfo{year}{2000}).

\bibitem[{\citenamefont{Billoire and Marinari}(2002)}]{BilloireMarinari02}
\bibinfo{author}{\bibfnamefont{A.}~\bibnamefont{Billoire}} \bibnamefont{and}
  \bibinfo{author}{\bibfnamefont{E.}~\bibnamefont{Marinari}}
  (\bibinfo{year}{2002}), \eprint{cond-mat/0202473}.

\bibitem[{\citenamefont{Fisher and Huse}(1986)}]{FisherHuse86}
\bibinfo{author}{\bibfnamefont{D.~S.} \bibnamefont{Fisher}} \bibnamefont{and}
  \bibinfo{author}{\bibfnamefont{D.~A.} \bibnamefont{Huse}},
  \bibinfo{journal}{Phys. Rev. Lett.} \textbf{\bibinfo{volume}{56}},
  \bibinfo{pages}{1601} (\bibinfo{year}{1986}).

\bibitem[{\citenamefont{Bray and Moore}(1987)}]{BrayMoore87}
\bibinfo{author}{\bibfnamefont{A.~J.} \bibnamefont{Bray}} \bibnamefont{and}
  \bibinfo{author}{\bibfnamefont{M.~A.} \bibnamefont{Moore}},
  \bibinfo{journal}{Phys. Rev. Lett.} \textbf{\bibinfo{volume}{58}},
  \bibinfo{pages}{57} (\bibinfo{year}{1987}).

\bibitem[{\citenamefont{Huse and Fisher}(1986)}]{HuseFisher86}
\bibinfo{author}{\bibfnamefont{D.~A.} \bibnamefont{Huse}} \bibnamefont{and}
  \bibinfo{author}{\bibfnamefont{D.~S.} \bibnamefont{Fisher}},
  \bibinfo{journal}{Phys. Rev. Lett.} \textbf{\bibinfo{volume}{57}},
  \bibinfo{pages}{2203} (\bibinfo{year}{1986}).

\bibitem[{\citenamefont{Banavar and Bray}(1987)}]{BanavarBray87}
\bibinfo{author}{\bibfnamefont{J.~R.} \bibnamefont{Banavar}} \bibnamefont{and}
  \bibinfo{author}{\bibfnamefont{A.~J.} \bibnamefont{Bray}},
  \bibinfo{journal}{Phys. Rev. B} \textbf{\bibinfo{volume}{35}},
  \bibinfo{pages}{8888} (\bibinfo{year}{1987}).

\bibitem[{\citenamefont{Nifle and Hilhorst}(1992)}]{NifleHilhorst92}
\bibinfo{author}{\bibfnamefont{M.}~\bibnamefont{Nifle}} \bibnamefont{and}
  \bibinfo{author}{\bibfnamefont{H.~J.} \bibnamefont{Hilhorst}},
  \bibinfo{journal}{Phys. Rev. Lett.} \textbf{\bibinfo{volume}{68}},
  \bibinfo{pages}{2992} (\bibinfo{year}{1992}).

\bibitem[{\citenamefont{Berker and Ostlund}(1979)}]{BerkerOstlund79}
\bibinfo{author}{\bibfnamefont{A.~N.} \bibnamefont{Berker}} \bibnamefont{and}
  \bibinfo{author}{\bibfnamefont{S.}~\bibnamefont{Ostlund}},
  \bibinfo{journal}{J. Phys. C} \textbf{\bibinfo{volume}{12}},
  \bibinfo{pages}{4961} (\bibinfo{year}{1979}).

\bibitem[{\citenamefont{Bray and Moore}(1985)}]{BrayMoore85}
\bibinfo{author}{\bibfnamefont{A.~J.} \bibnamefont{Bray}} \bibnamefont{and}
  \bibinfo{author}{\bibfnamefont{M.~A.} \bibnamefont{Moore}},
  \bibinfo{journal}{Phys. Rev. B} \textbf{\bibinfo{volume}{31}},
  \bibinfo{pages}{631} (\bibinfo{year}{1985}).

\bibitem[{\citenamefont{Palassini and Young}(1999)}]{PalassiniYoung99}
\bibinfo{author}{\bibfnamefont{M.}~\bibnamefont{Palassini}} \bibnamefont{and}
  \bibinfo{author}{\bibfnamefont{A.~P.} \bibnamefont{Young}},
  \bibinfo{journal}{Phys. Rev. Lett.} \textbf{\bibinfo{volume}{83}},
  \bibinfo{pages}{5126} (\bibinfo{year}{1999}).

\end{thebibliography}

\end{document}